# FAST BEAM STACKING USING RF BARRIERS*

W. Chou#, D. Capista, J. Griffin, K-Y. Ng and D. Wildman, Fermilab, Batavia, Illinois, USA

*Abstract*

Two barrier RF systems were fabricated, tested and installed in the Fermilab Main Injector. Each can provide 8 kV rectangular pulses (the RF barriers) at 90 kHz. When a stationary barrier is combined with a moving barrier, injected beams from the Booster can be continuously deflected, folded and stacked in the Main Injector, which leads to doubling of the beam intensity. This paper gives a report on the beam experiment using this novel technology.

## INTRODUCTION

At present the Fermilab Main Injector (MI) can deliver about $3.3 \times 10^{13}$ protons at 120 GeV to the targets every 2.4 seconds for a beam power of about 270 kW. The limit mainly comes from two sides: the injection time (67 ms per Booster pulse) and the number of Booster pulses that can be injected into the MI. The recently approved NOvA project would convert the Recycler to a proton accumulator and thus eliminate the long injection front porch in the MI. In order to increase the number of Booster pulses, the slip stacking method has been studied and made significant progress [1]. This paper introduces another stacking method that makes use of RF barriers.

A new type of barrier RF system was discussed in previous publications [2-4]. It is made of an RF cavity and a modulator. Both are connected by two impedance matching transformers and long 50 Ω cables. The cavity is made of Finemet cores, a low Q high μ material manufactured by Hitachi Co. The modulator uses a pair of bipolar high voltage fast MOSFET switches made by Behlke Co. The first unit was installed in the MI in 2003. After successful operations, a second unit was installed in 2006. Figure 1 shows the two RF cavities and modulators.

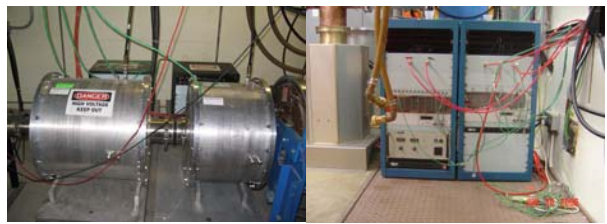

Figure 1: Left – RF barrier cavities, right - modulators.

By using two RF barriers, one can perform continuous beam stacking to double the beam intensity in the MI by doubling the number of Booster injections. It is called *fast stacking* to distinguish from the *slow stacking* that uses adiabatic compression as reported previously [3]. This novel method was first suggested by J. Griffin [5]. It can inject 12 Booster pulses to the MI instead of 6 or 7 as in today's operation.

## METHOD

The injected Booster beam is intentionally given a momentum offset Δp, which leads to a drift of the beam at the rate:

$$\frac{\Delta t}{T} = -\frac{\Delta f_0}{f_0} = \eta \frac{\Delta p}{p}$$

in which $T$ is the revolution time (11.2 μs), $f_0$ the revolution frequency (89.815 kHz), $p$ the normal injection momentum (8.89 GeV/c), and $\eta$ the slip factor ($-8.88 \times 10^{-3}$). The offset is chosen such that the beam would drift half the pulse width or 0.8 μs during one Booster cycle (67 ms). So the spacing between two consecutive Booster pulses is 0.8 μs instead of the normal value of 1.6 μs. This makes it possible to inject 12 pulses instead of 6 for the same total beam width of 9.6 μs, leaving a room (1.6 μs) for the kicker firing time. A stationary RF barrier prevents the earlier injected beam from entering the space reserved for later pulses. A moving barrier serves the purpose for reducing the momentum spread of the stacked beam. The method is illustrated in Figure 2.

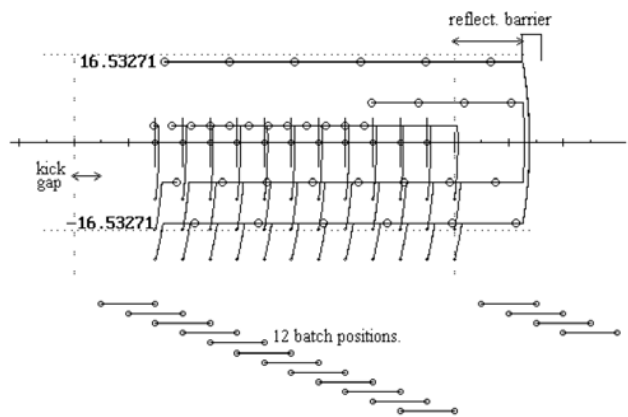

Figure 2: Injection of 12 Booster pulses using RF barriers.

There are two conceivable ways to create an injection momentum offset. One is to raise the injection magnetic field $B$, another to increase the injection RF frequency $f$ (which is locked between the MI and Booster at injection). However, the former is wrong because it would put the beam on a smaller closed orbit that makes it impossible to maintain the right harmonic number ($h = 588$). Therefore, it was decided in our experiment to raise the injection RF frequency while keeping $B$ at the normal value.

Because during MI injection the slip factor has opposite sign in the MI ($-8.88 \times 10^{-3}$) and in the Booster ($2.27 \times 10^{-2}$), a higher RF frequency would have opposite effect on the beam energy in the two machines. The net change

___________________________________________
* Work supported by the U.S. Department of Energy under Contract No. DE-AC02-07CH11359.
#chou@fnal.gov

is a sum of the two effects. The following is a numerical illustration.

The normal RF frequency at injection is 52,811,400 Hz. To lower the injection beam energy, the frequency was changed to 52,812,060 Hz, an increase of $\Delta f = 660$ Hz. This corresponds to an increase of momentum in the MI:

$$\frac{\Delta p}{p} = \frac{-\Delta f / f}{\eta} = +0.141\%$$

In the Booster, however, this frequency corresponds to a momentum decrease since $\eta$ has positive sign (above transition):

$$\frac{\Delta p}{p} = \frac{-\Delta f / f}{\eta} = -0.055\%$$

Therefore the momentum offset of the Booster beam in the MI is a sum of the two, i.e. $\Delta p/p = -0.196\%$, or an energy offset $\Delta E = -18$ MeV.

The measured energy spread of the Booster beam at low intensity is ±6.1 MeV (see below). So the highest energy particles in the beam have an offset of –12 MeV, which gives a drift:

$$\Delta t = \eta \frac{\Delta p}{p} T = 0.8 \text{ μs} \text{ per Booster cycle}$$

This rate meets the requirement.

## PROCEDURE

At injection, the RF system in the MI was phase locked to the Booster RF at 52,812,060 Hz and paraphrased to zero amplitude. So the injected beam was debunched and had a momentum offset of –0.196%.

Two RF barriers, each of ±8 kV height and 400 ns width, were turned on from the beginning until the end of injection. One was stationary serving as a reflection wall. Another was moving at a speed of 0.8 μs every Booster cycle as shown in Figure 3.

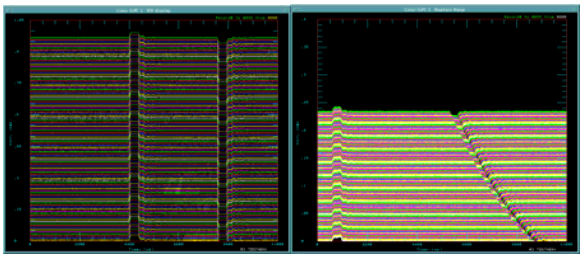

Figure 3: Left – stationary barrier, right – moving barrier.

After the first Booster pulse entered the MI, it drifted about 0.8 μs before the second pulse was injected. The spacing between two consecutive pulses was 0.8 μs, or 42 RF buckets. After a total of 12 pulses were injected, the main RF system was gradually turned on to adiabatically recapture the beam into the 53 MHz buckets while the barrier RF was turned off. It was then followed by acceleration to 120 GeV.

## EXPERIMENTAL RESULTS

Because this method has strong dependence on the momentum spread of the incoming beam, we started with low intensity beam (1-turn injection from the linac to Booster). Figure 4 is a beam test of stationary barriers. Two consecutive Booster pulses were injected. Due to momentum offset they started to drift to the right but were stopped by the barrier and reflected back to the left. When they met another barrier on the left, they were stopped again and contained within two barriers.

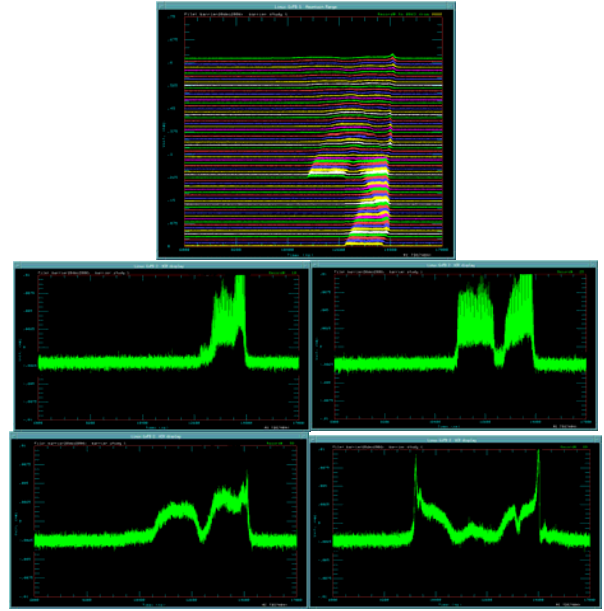

Figure 4: Top – two consecutive pulses in MI (Mountain View plot), middle left – 1st pulse, middle right – two pulses, bottom left – beam stopped by right barrier, bottom right – beam confined within two barriers.

In normal operation, 6 Booster pulses have a width of 9.6 μs, or 6/7 of the MI ring size. In this experiment, however, the width of 6 stacked pulses was only 5.5 μs, or less than half of the MI size as shown in Figure 5.

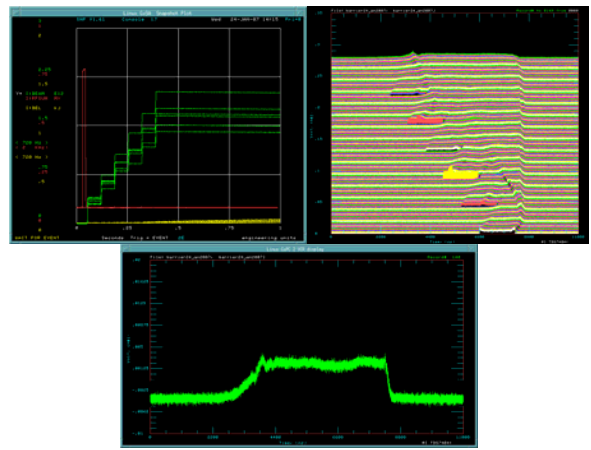

Figure 5: Top left – 6 injections (green lines), top right – 6 pulses in Mountain View plot, bottom – the total width of 6 pulses was 5.5 μs.

Figure 6 shows the injection of 8, 10 and 12 pulses respectively.

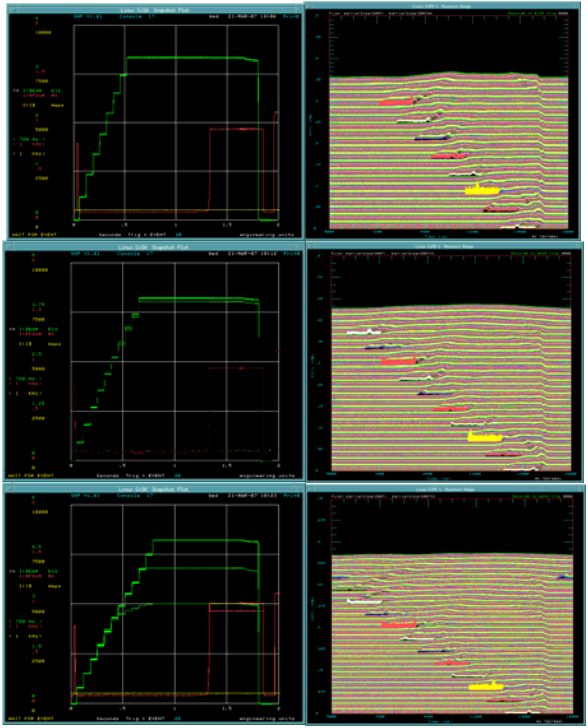

Figure 6: Top left – 8 injections (green lines), top right – 8 pulses in Mountain View plot, middle – 10-pulse injection, bottom – 12-pulse injection.

It is interesting to compare bottom right of Figure 6 to Figure 2. The 12 colored "boats" mimic the 12 batches in the simulation. Another interesting thing is in bottom left of Figure 6, where the beam stacking stopped when the barriers were turned off.

Figure 7 shows the recaptured beam (top) and the beam accelerated to 120 GeV (bottom).

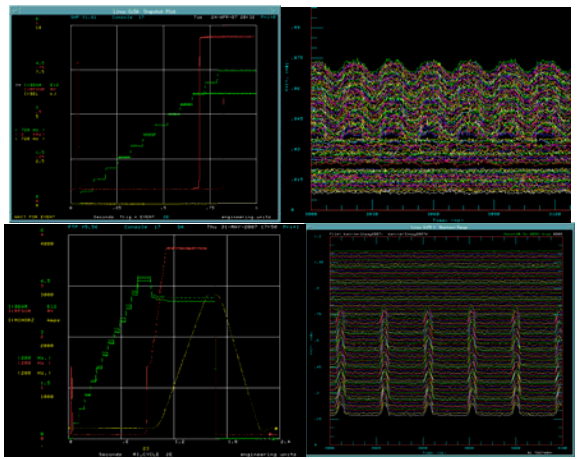

Figure 7: Top left – 12 injections (green lines), red line is the RF voltage, top right – recaptured beam at 8 GeV, bottom left – 12 injected pulses accelerated to 120 GeV, bottom right – beam at 120 GeV.

During injection and recapture, there was no noticeable beam loss. However, there were losses at the beginning of acceleration as shown in the bottom left of Figure 7. This was mainly due to dc beam that was not captured by the 53 MHz RF bucket. Simulation indicates this loss could be eliminated by improving the RF recapture procedure.

## DIRECT MEASUREMENT OF BEAM MOMENTUM SPREAD

A "bonus" of this experiment is a new method to measure the beam momentum spread directly rather than using a calculation based on beam width measurement and known RF voltage. This new method makes use of the fact that a beam will be completely contained by a barrier unless its momentum spread is larger than the barrier height. The barrier height is well known:

$$\Delta E = \beta \sqrt{\frac{2eV_b E}{\eta} \times \frac{T_b}{T}}$$

in which $\beta$ is the relativistic factor, $V_b$ and $T_b$ the barrier voltage and width, respectively, and $E$ the beam energy. Figure 8 shows the beam started to penetrate when the barrier was lowered from 9.5 kV to 9 kV, which gave a beam energy spread of ±6.1 MeV

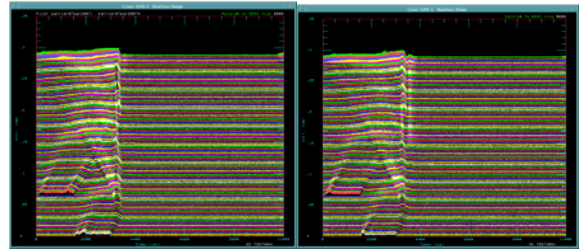

Figure 8: Left – no penetration at 9.5 kV, right – penetration at 9 kV.